\documentclass[%
pra,reprint,superscriptaddress,notitlepage,
showpacs,
amsmath,
amssymb,
]{revtex4-1}

\usepackage{graphicx}
\usepackage{dcolumn}
\usepackage{bm}

\usepackage[utf8]{inputenc} 
\usepackage{mathrsfs}
\usepackage{xcolor}
\usepackage{amsmath,amsfonts}
\usepackage{indentfirst}
\usepackage{natbib}

\newcommand{\varzero}{\alpha}
\newcommand{\varone}{A^1}
\newcommand{\vartwo}{A^2}

\begin{document}

\title{Quantum Walks and discrete Gauge Theories}

\author{Pablo Arnault}
\email{pablo-arnault@hotmail.fr}
\affiliation{LERMA, UMR 8112, UPMC and Observatoire de Paris, 61 Avenue de l'Observatoire, 75014 Paris, France}

\author{Fabrice Debbasch}
\email{fabrice.debbasch@gmail.com}
\affiliation{LERMA, UMR 8112, UPMC and Observatoire de Paris, 61 Avenue de l'Observatoire, 75014 Paris, France}

\date{\today}

\begin{abstract}
A particular example is produced to prove that quantum walks can be used to simulate full-fledged discrete gauge theories. 
A new family of $2D$ walks is introduced and its continuous limit is shown to coincide with the dynamics of a Dirac fermion coupled to arbitrary electromagnetic fields. The electromagnetic interpretation is extended beyond the continuous limit by proving that these DTQWs exhibit an exact discrete local $U(1)$ gauge invariance and possess a discrete gauge-invariant conserved current. A discrete gauge-invariant electromagnetic field is also constructed and that field is coupled to the conserved current by a discrete generalization of Maxwell equations. The dynamics of the DTQWs under crossed electric and magnetic fields is finally explored outside the continuous limit by numerical simulations. Bloch oscillations and the so-called ${\bf E} \times {\bf B}$ drift are recovered in the weak-field limit. Localization is observed for some values of the gauge fields.

\end{abstract}


\maketitle


\section{Introduction} 

Discrete Time Quantum Walks (DTQWs) are formal generalizations of classical random walks. They were first studied in a systematic fashion by D. A. Meyer \cite{Meyer96a}, while similar but different quantum discrete dynamics was first considered in
\cite{FeynHibbs65a, Grossing_Zeilinger_98, ADZ93a}.
DTQWs have been realized experimentally  with a wide range of physical objects and setups \cite{Schmitz09a, Zahring10a, Schreiber10a, Karski09a, Sansoni11a, Sanders03a, Perets08a}, and are studied in a large variety of contexts, ranging from fundamental quantum physics \cite{Perets08a, var96a} to quantum algorithmics \cite{Amb07a, MNRS07a}, solid state physics \cite{Aslangul05a, Bose03a, Burg06a, Bose07a} and biophysics \cite{Collini10a, Engel07a}. 

Particular quantum cellular automata \cite{Succi93, Birula94}, among which DTQWs \cite{Arrighi14} defined on various regular lattices \cite{Chandrashekar13}, are known to reproduce, in the continuous limit, the dynamics of free Dirac fermions in one, two or three spatial dimensions.
Recently, such connexions have been extensively extended to Dirac fermions coupled to gauge fields \cite{DDMEF12a, mesch13a, ced13,DMD13b, DMD14, SFGP15a, Arrighi_curved_1D_15, ADmag16}.
More precisely, $1D$ DTQWs have been proposed which reproduce the dynamics of Dirac fermions coupled to arbitrary electric \cite{DDMEF12a, mesch13a, ced13} and/or gravitational \cite{DMD13b, DMD14, SFGP15a, Arrighi_curved_1D_15} fields, and a $2D$ DTQW simulating the coupling of a Dirac fermion to a constant uniform magnetic field has been proposed in \cite{ADmag16}. 
 

For all existing DTQWs, the gauge fields are encoded in the time and space dependence of the operator advancing the fermion in discrete spacetime. 
They act on the fermion but the dynamics of the fermion has no effect on the gauge fields. In other words, the gauge fields play the roles of imposed external fields. In particular, they are not advanced by their own discrete dynamical equations, as is for example the case in Lattice Gauge Theories (LGTs). The main purpose of this article is to remedy this problem and introduce the first complete self-consistent model based on DTQWs where both the fermions and the fields are advanced by compatible discrete dynamical equations.

The first brick in such a self-consistent model is a family of DTQWs which exhibit an exact discrete gauge invariance associated to a certain group $G$ and which describes the coupling of Dirac fermions to arbitrary $G$-gauge fields. The existing literature contains only one such family. The associated gauge group is $U(1)$ and the DTQWs describe the coupling of $1D$ Dirac fermions to arbitrary electric fields. Electromagnetism is however degenerate in $1D$. There is no magnetic field and Maxwell equations reduce to the Maxwell-Gauss equation, which contains no time derivative. We therefore switch to $2D$ and introduce
%
a new family of DTQWs 
whose continuous limit coincides with the dynamics of a Dirac fermion coupled to $2D$ arbitrary electromagnetic fields. We then show that these
DTQWs admit (i) an exact discrete $U(1)$ gauge invariance (ii) a gauge-invariant discrete electromagnetic tensor ({\sl i.e.} gauge invariant electric and magnetic fields defined on the discrete lattice of the DTQWs) (iii) a discrete conserved current. We finally combine the discrete electromagnetic tensor and the discrete conserved current into discrete gauge-invariant Maxwell equations which imply current conservation. 
This literal material is complemented by numerical computations which explore how the DTQWs which serve as a basis for the whole construct behave outside the continuous limit. Even outside this limit, 
the DTQWs display in the weak field regime several well-known features usually associated to standard continuous motions in electromagnetic fields, including Bloch oscillations and the so-called $\mathbf{E} \times \mathbf{B}$ drift. In the regime of strong fields, the discrete dynamics depends crucially on whether the fields are rational or not, as expected from previous work on other DTQWs \cite{ced13, mesch13a, Yalcinkaya15}.
We finally discuss possible applications of our results to quantum simulation and quantum algorithmics and highlight how the new discrete gauge theory based on DTQWs differs from standard Lattice Gauge Theories (LGTs).

\section{Electromagnetic DTQWs} 
\subsection{The walks and their formal continuous limit}

We consider DTQWs with two-component wavefunctions ($2D$ coin or spin space) defined on a discrete $(1 + 2)$-dimensional spacetime where instants are labeled by the  index $j \in {\mathbb N}$ and space points on the $2D$ square lattice are labeled by the indices $(p, q) \in {\mathbb Z}^2$. The evolution equation for the wave-function reads
\begin{align} \label{eq:fund}
\Psi_{j + 1, p, q} & =  \mathbf{U} \! \left(\theta^-(\epsilon_m,m),\epsilon_A \vartwo_{j, p, q}, \epsilon_A A^0_{j, p, q}\right)\mathbf{T}_2 \\ \nonumber
& \ \ \ \ \times \mathbf{U} \! \left(\theta^+(\epsilon_m,m),\epsilon_A \varone_{j, p, q},0 \right) \mathbf{T}_1 \, \Psi_{j, p, q} \ ,
\end{align}
where the action of the shift operators $\mathbf{T}_1$ and $\mathbf{T}_2$
on the $2D$ wave-function $\Psi_{j, p, q} =  \left(\psi^-_{j, p, q}, \psi^+_{j, p, q} \right)^{\top}$ (the superscript $\top$ denotes the transposition) is:
\begin{eqnarray}
\mathbf{T}_1 \Psi_{j, p, q} & = & \left( \psi^-_{j, p+1, q}, \psi^+_{j, p-1, q} \right)^{\top} \\
\mathbf{T}_2 \Psi_{j, p, q} & = & \left( \psi^-_{j, p, q+1}, \psi^+_{j, p, q-1} \right)^{\top} . \nonumber
\end{eqnarray}
%
The coin operator $\mathbf{U}(\theta, \xi,\varzero) \in U(2)$ is the product of three simpler operators:
\begin{align}
\mathbf{U}(\theta, \xi,\varzero) &= e^{i\alpha} \mathbf{1} \times \mathbf{C}(\theta) \times \mathbf{S}(\xi) \\
&=
\left[
  \begin{array}{c c}
   e^{i\alpha} &
   0 \\
   0 &
   e^{i\alpha}
  \end{array}
  \right]
\left[
  \begin{array}{c c}
  \cos \theta &
   i\sin \theta \\
   i\sin \theta &
  \cos \theta
  \end{array}
  \right]
\left[
  \begin{array}{c c}
   e^{i\xi} &
   0 \\
   0 &
   e^{-i\xi}
  \end{array}
  \right]. \nonumber
\end{align}
The first operator $\mathbf{S}(\xi)$ is a spin-dependent phase shift parametrized by the angle $\xi$, the second operator $\mathbf{C}(\theta)$ is a standard coin operator with angle $\theta$
and the third operator performs a global multiplication by the phase~$\alpha$.

In the continuous limit, the parameter $m$ which enters the definition of the constant angles $\theta^{\pm} (\epsilon_m, m) = \pm \frac{\pi}{4} - \epsilon_m \, \frac{m}{2}$, will be interpreted as the mass of the walk and the three angles $A^0$, $A^1$,  $A^2$, which may depend on $(j,p,q)$, will be interpreted as the components of an electromagnetic potential. The positive parameters $\epsilon_m$ and $\epsilon_A$ are introduced to trace the importance of $m$ and $A$ and will tend to zero in the continuous limit. All the parameters and angles are dimensionless.

The formal continuous limit of the DTQWs (\ref{eq:fund}) can be determined by the method used in  \cite{ADmag16, DMD14, DMD13b, DMD12a, DMDmag13, DMR97a}: we first introduce a (dimensionless) spacetime-lattice step $\epsilon_l$ 
and interpret any $(j, p, q)$-dependent quantity $Q_{j, p, q}$ as the value taken by a function $Q(X^0, X^1, X^2)$ at time 
$X^0_j = j \epsilon_l$ and spatial position $(X^1_p = p \epsilon_l$, $X^2_q = q \epsilon_l)$. We then consider the scaling $\epsilon_m = \epsilon_A = \epsilon_l = \epsilon$ and let $\epsilon$ tend to zero.
We expand Eq. (\ref{eq:fund}) at first order in $\epsilon$ around the generic spacetime point $(X_j^0,X_p^1,X_q^2)$. 
For the continuous limit to exist, the zeroth-order terms of the expansion must balance each other; this contraint is automatically verified by the DTQWs defined by (\ref{eq:fund}).

The first-order terms of the expansion in $\epsilon$ deliver the 
differential equation which determines the dynamics of the walker in the continuous limit. This equation reads:
\begin{equation} \label{eq:Dirac}
(i\gamma^{\mu} {\mathcal D}_{\mu} - m)\Psi = 0 \ .  
\end{equation}
Here, the $\gamma$ matrices are defined by $\gamma^0 = \sigma_1$, $\gamma^1 = i \sigma_2$, $\gamma^2 = i \sigma_3$, where the Pauli matrices are
\begin{equation}
\sigma_1 = \left(
    \begin{array}{cc}
       0 & 1 \\
       1 & 0 
    \end{array}
  \right), \ \  \sigma_2 = \left(
    \begin{array}{cc}
       0 & -i \\
       i & 0 
    \end{array}
  \right),  \ \  \sigma_3 =  \left(
    \begin{array}{cc}
       1 & 0 \\
       0 & -1 
    \end{array}
  \right) ,
\end{equation} 
\noindent
and ${\mathcal D}_\mu = \partial_\mu - i A_\mu$ is the covariant derivative associated to Maxwell electromagnetism, with $A_0 = A^0$, $A_1 = - A^1$, $A_2 = - A^2$. The $\gamma$ matrices satisfy the $(1+ 2)$-dimensional flat-spacetime Clifford algebra $\gamma^{\mu} \gamma^{\nu} + \gamma^{\nu} \gamma^{\mu} = 2 \eta^{\mu\nu } \mathbf{1}$, where $ [ \eta^{\mu\nu} ]  = \mathrm{diag}(1,-1,-1) $  is the Minkowskian metric.
Equation (\ref{eq:Dirac}) is the Dirac equation describing the dynamics of a spin $1/2$ fermion of mass $m$ and charge $-1$ coupled to the electromagnetic potential $A$. To consider a generic charge $g$, just perform the substitution $A \rightarrow -gA$. In Eq. (\ref{eq:Dirac}), the characteristic speed is $1$ because we have chosen the same value $\epsilon_l$ for the dimensionless time and space steps.



\subsection{Rate of convergence towards the continuous limit}

\begin{figure}[!h]
\vspace{0.2cm} 
\centering
\includegraphics[width=8.5cm]{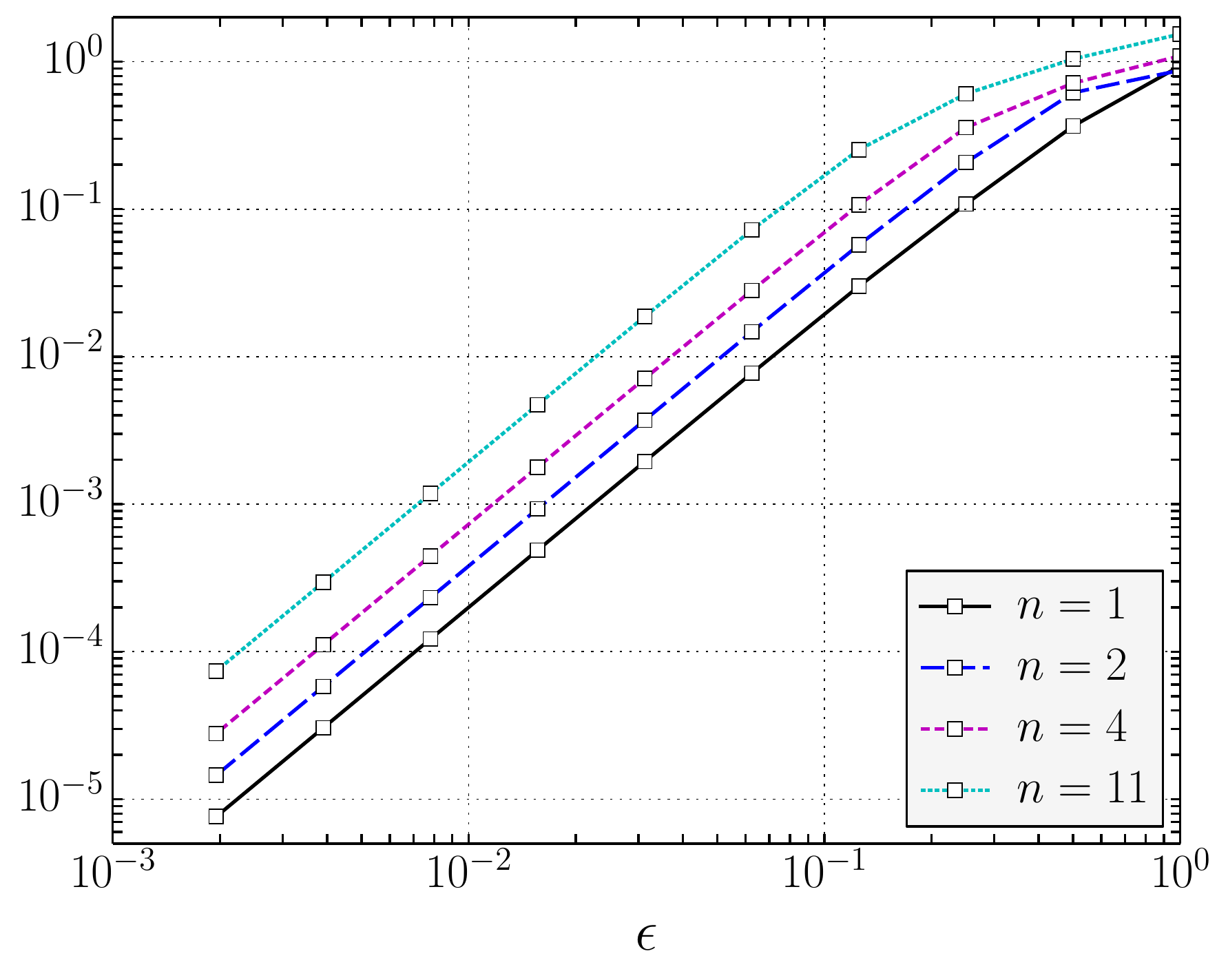}
\vspace{-0.6cm}
\caption{\small Distance $\delta_l(K = 0)$ as a function of $\epsilon$ for various values of $l = (+,n)$.}
\label{varying_level}
\end{figure}

\begin{figure}[!h]
\centering
\includegraphics[width=8.5cm]{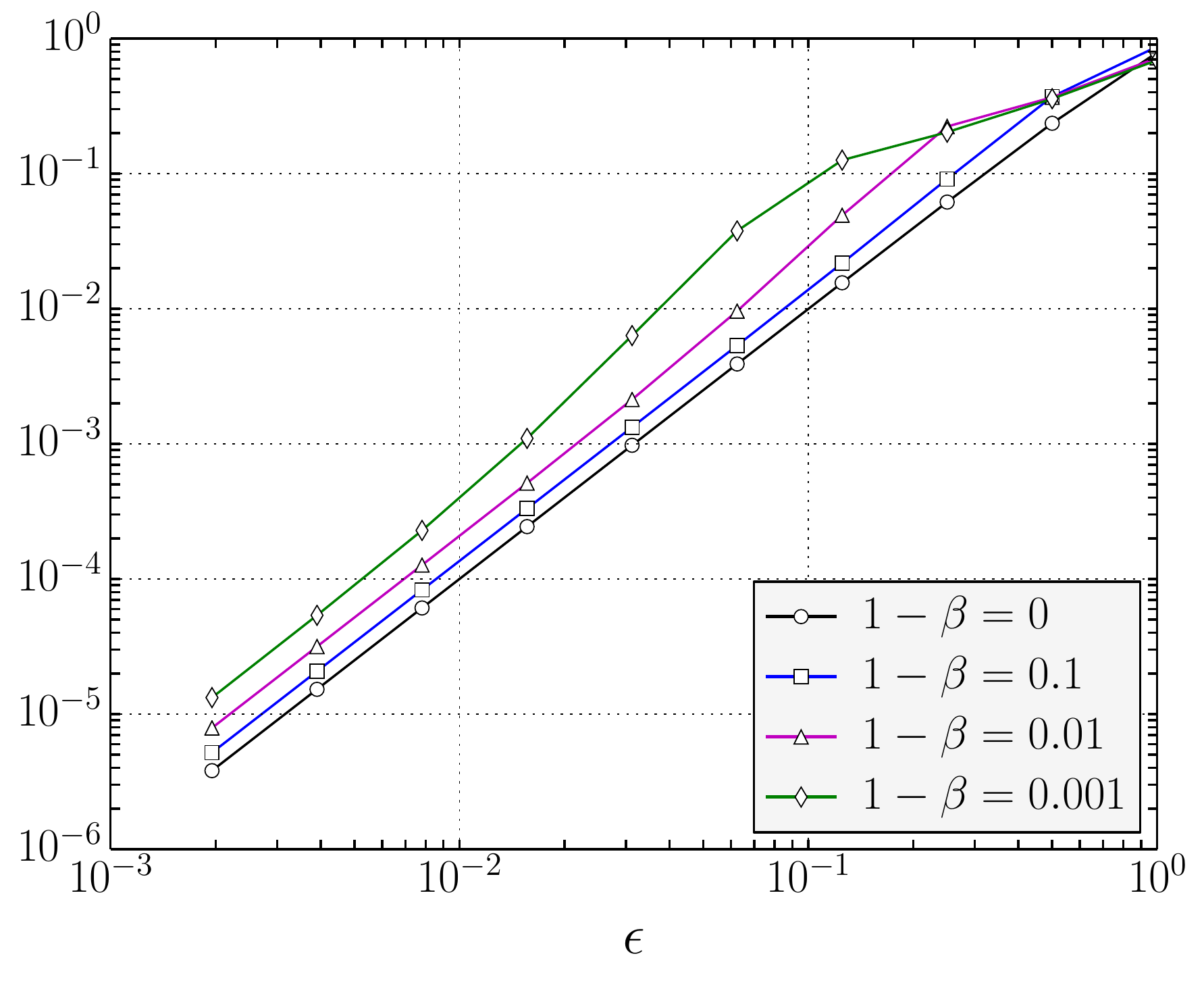}
\vspace{-0.6cm}
\caption{\small Distance $\delta_l(K = 0)$ as a function of $\epsilon$ for various values of $\beta$.}
\label{varying_beta}
\end{figure}

Since the formal continuous limit is obtained from the DTQW dynamics by keeping terms that are first order in $\epsilon$, one expects the discrepancy between a solution of the DTQW and the corresponding solution of the Dirac equation to scale as $\epsilon^2$. This
can be tested numerically by computing the distance between an exact time-independent solution of the Dirac equation and the time-evolution of this solution by the DTQW. 
For simplicity we choose $A_0 = - E X^1, A_1 = 0, A_2 = -B X^1$, which does not depend on $X^2$ and generates crossed, constant and uniform electric and magnetic fields.
The Hamiltonian and the momentum in the $X^2$ direction can then be diagonalized simultaneously. For $E = 0$, the eigenstates are called relativistic Landau levels  \cite{ADmag16}. For $0 < \beta = E/B < 1$, the eigenstates can be obtained from the relativistic Landau levels by a boost of velocity $\beta$. The resulting eigenstates $\phi_{l, K} (X^1, X^2) = \Phi_l(X^1,K) \exp(i K X^2)$ and eigen-energies $\mathcal{E}_{l,K}$
are labelled by a couple $(l, K)$ where $l = 0$ or $l = (\pm, n)$ with $n\in \mathbb{N^{\ast}}$ and $K$ is the eigen-momentum in the $X^2$ direction. 
In one time step $\epsilon$, the DTQW evolves $\Phi_{l}(X^1, K)$ into a certain function $W_l(X^1, K)$ which should be approximated, at first order in $\epsilon$, by ${\tilde W}_l(X^1, K)= \exp(-i \mathcal{E}_{l,K} \times \epsilon) \Phi_{l}(X^1,K)$. For each $K$, the distance between the two functions
 ${W}_l (\cdot , K)$ and 
${\tilde W}_l(\cdot , K)$, can be evaluated by
\begin{equation} \label{eq:relative difference}
\delta_l(K) \equiv \frac{\| W_l(\cdot, K) - {\tilde W}_l(\cdot, K) \|}{\| {\tilde W}_l(\cdot, K) \|} \ ,
\end{equation}
where $\| \cdot \|$ stands for the $L^2$ norm of a position ($X^1$-)dependent function $\Psi$ defined on the lattice:
\begin{equation} \label{eq:L2_norm}
\| \Psi \| = \left[ \sum\limits_{\substack{p = - p_{\mathrm{max}}(\epsilon)}}^{p_{\mathrm{max}}(\epsilon)}  \left( |\psi^{-}(X^1_p)|^2 + |\psi^{+}(X^1_p)|^2 \right)  \epsilon \right]^{\frac{1}{2}} ,
\end{equation}
where $p_{\mathrm{max}}(\epsilon)$ scales as $1/\epsilon$.
Figures \ref{varying_level} and \ref{varying_beta} display how $\delta_l(K = 0)$ scales with $\epsilon$ for various values of $l = (+,n)$ and for various values of $\beta$, having fixed $B=1$ and $m=1$. These figures clearly confirm that $\delta_l(K = 0)$ scales as $\epsilon^2$ for a large range of $\epsilon$-values

\section{Discrete gauge invariance and electromagnetic fields}

 The discrete equations (\ref{eq:fund}) are invariant, not only under a global phase-change of the spinor $\Psi$, but also under the more general, local gauge transformation 
\begin{eqnarray}
\Psi_{j, p, q} & \rightarrow & \Psi'_{j, p, q} = e^{-i \phi_{j, p, q}} \ \! \Psi_{j, p, q} 
\\
(A_\mu)_{j, p, q} & \rightarrow & (A'_\mu)_{j, p, q} = (A_\mu)_{j, p, q} - \left(d_\mu \phi \right)_{j, p, q} \nonumber, 
\end{eqnarray}
where the three `discrete-derivative' (finite-difference) operators $d_\mu$ are defined by 
\begin{equation}
d_0  = (L - \Sigma_2 \Sigma_1)/\epsilon_A
, \ \ \ \ 
d_1  = \Delta_1/\epsilon_A
, \ \ \ \ 
d_2  = \Delta_2 \Sigma_1/\epsilon_A
, \
\end{equation}
with 
\begin{align} \label{eq:operators}
(L Q)_{j, p, q} &= Q_{j+1, p, q} \nonumber \\
(\Sigma_1 Q)_{j, p, q} &= \left(Q_{j, p+1, q} + Q_{j, p-1, q}\right)/2 \nonumber \\
(\Sigma_2 Q)_{j, p, q} &= \left(Q_{j, p, q+1} + Q_{j, p, q-1}\right)/2 \\
(\Delta_1 Q)_{j, p, q} &= \left(Q_{j, p+1, q} - Q_{j, p-1, q}\right)/2 \nonumber \\
(\Delta_2 Q)_{j, p, q} &= \left(Q_{j, p, q+1} - Q_{j, p, q-1}\right)/2 \ . \nonumber
\end{align}

This local gauge invariance is a discrete version of the standard continuous $U(1)$ local gauge invariance associated to electromagnetism and displayed by the Dirac equation (\ref{eq:Dirac}). A straightforward computation now shows that the three quantities $F_{01}$, $F_{02}$ and $F_{12}$ defined by
\begin{equation}
(F_{\mu \nu})_{j, p, q}  = \left(d_\mu  A_\nu\right)_{j, p, q}  - \left(d_\nu  A_\mu\right)_{j, p, q}
\end{equation}
are gauge invariant. These are clearly discrete versions of the usual electromagnetic tensor components $F_{\mu \nu} = \partial_\mu A_\nu - \partial_\nu A_\mu$. In particular, $F_{01}$ and $F_{02}$ represent respectively the two components $E_1$ and $E_2$ of a $2D$ discrete electric field (parallel to the plan of the $(p, q)$-grid) and the component $F_{12}$ represents a discrete magnetic field $B_3$ perpendicular to the plan of the $(p, q)$-grid. 

\section{Gauge-invariant conserved current and discrete Maxwell equations}

Let 
\begin{equation}
{\widetilde \Psi}_{j, p, q} =  \mathbf{U} \! \left(\theta^+(\epsilon_m,m),\epsilon_A \varone_{j, p, q},0 \right) \mathbf{T}_1 \, \Psi_{j, p, q} 
\label{eq:psi_tilde}
\end{equation}
be the state of the walker after the shift along the $p$ direction and the first coin operation.
The spatial density associated to ${\widetilde \Psi}_{j, p, q}$ is
\begin{align}
{\widetilde \Psi}^{\dag}_{j, p, q} {\widetilde \Psi}_{j, p, q} &= | {\Psi}^-_{j,p+1,q} |^2 + |{\Psi}^+_{j,p-1,q}|^2 \ .
\end{align}
Introducing notations $\rho = |{\Psi}^-|^2 + |{\Psi}^+|^2$ and $\mathcal{J}=| {\Psi}^+ |^2 - |{\Psi}^-|^2$, we can rewrite the previous equation as
\begin{equation} \label{eq:first_half_time_step}
{\widetilde \rho}_{j, p, q} = (\Sigma_1 \rho)_{j, p, q} - (\Delta_1 \mathcal{J})_{j, p, q} \ .
\end{equation}
The same computation carried out for $\Psi_{j + 1, p, q}  =  \mathbf{U} \! \left(\theta^-(\epsilon_m,m),\epsilon_A \vartwo_{j, p, q}, \epsilon_A A^0_{j, p, q}\right)\mathbf{T}_2 \, {\widetilde \Psi}_{j, p, q}$ results in
\begin{equation} \label{eq:second_half_time_step}
(L \rho)_{j, p, q} = (\Sigma_2 \widetilde \rho)_{j, p, q}  - (\Delta_2 \mathcal{\widetilde J})_{j, p, q} \ .
\end{equation}
Inserting (\ref{eq:first_half_time_step}) into (\ref{eq:second_half_time_step}) gives the discrete conservation equation
\begin{equation}
(L \rho)_{j, p, q} = (\Sigma_2 \Sigma_1 \rho)_{j, p, q} - (\Sigma_2 \Delta_1 \mathcal{J})_{j, p, q} - (\Delta_2 \mathcal{\widetilde J})_{j, p, q}\ .
\end{equation}

All operators defined in (\ref{eq:operators}) commute with each other; in particular, $\Sigma_2 \Delta_1 = \Delta_1 \Sigma_2$, so that the conservation equation can be written
\begin{equation} \label{eq:discrete_conservation_eq}
(D_\mu J^\mu)_{j, p, q} = 0 \ ,
 \end{equation}
where the new finite-difference operators $D_\mu$ read,
\begin{equation}
D_0 = d_0, \ \ \ \ \ D_1 = d_1 \Sigma_2, \ \ \ \ \  D_2 = \Delta_2/\epsilon_A,
\end{equation}
and the probability current on the square lattice is given by
\begin{align}
J^0_{j, p, q} & =  \rho_{j,p,q} = \ \mid \psi^+_{j, p, q} \mid^2 +  \mid \psi^-_{j, p, q} \mid^2 \nonumber \\
J^1_{j, p, q} & =  \mathcal{J}_{j,p,q} = \ \mid \psi^+_{j, p, q} \mid^2 -  \mid \psi^-_{j, p, q} \mid^2  \\
J^2_{j, p, q} & =  \mathcal{\widetilde{J}}_{j,p,q} = \ \mid {\widetilde \psi}^+_{j, p, q} \mid^2 -  \mid {\widetilde \psi}^-_{j, p, q} \mid^2 \ .\nonumber
\end{align}
In the continuous limit, the discrete conservation equation (\ref{eq:discrete_conservation_eq}) becomes the standard conservation equation $\partial_\mu j^\mu = 0$ of the $2D$ Dirac current $j^\mu = {\bar \Psi} \gamma^\mu \Psi$, with ${\bar \Psi} =  \Psi^{\dag} \gamma^0$.

Having identified the discrete current $J^\mu$ and the finite-difference operators involved in the discrete continuity equation (\ref{eq:discrete_conservation_eq})  makes it possible to write the following simple discrete equivalent to Maxwell equations
\begin{equation} \label{eq:discrete_Maxwell}
(D_{\mu} F^{\mu \nu})_{j, p, q} = (J^{\nu})_{j, p, q} \ ,
\end{equation}
which connects the discrete electromagnetic tensor $(F_{\mu \nu})_{j,p,q}$ to the discrete current $(J^\mu)_{j,p,q}$. Indeed, Eq. (\ref{eq:discrete_Maxwell}) has the standard Maxwell equations as continuous limit, and ensures the conservation of the discrete current $(J^{\mu})_{j,p,q}$ because it implies $(D_\nu J^{\nu})_{j, p, q} = (D_\nu D_\mu F^{\mu \nu})_{j, p, q}$, which vanishes identically because operators $D_\mu$ commute with each other and because $(F^{\mu \nu})_{j,p,q}$ is antisymmetric.

\section{Simulations outside the continuous limit} 

We now focus on constant and uniform discrete electric and magnetic fields, 
for example ${\bf E} = E {\bf u}_1$ and ${\bf B} = B {\bf u}_3$ where ${\bf u}_1$ and ${\bf u}_3$ are two unitary vectors respectively along the $p$- (or $X^1$-)axis of the grid and perpendicular to the plane of the grid. A potential generating these fields is
$(A_0)_{j, p, q}  =  - E  p \epsilon_l$, $(A_1)_{j, p, q} =  0$,  $(A_2)_{j, p, q}  =  -B  p \epsilon_l$.
Walks with $B = 0$ ({\sl resp.} $E = 0$) will be referred to as $E$-walks ({\sl resp.} $B$-walks). Walks with $E \ne 0$ and $B \neq 0$ will be referred to as $EB$-walks. 

Quantities of particular interest are the probability of presence of the walker $P_{j, p, q} =  |\psi^-_{j, p, q}|^2 + |\psi^+_{j, p, q}|^2$ and, for $l=p$ or $q$, its time-dependent $l$-mean ({\sl resp.} $l$-spread), defined as the time-dependent average ({\sl resp.} square-rooted average) value of $l$ ({\sl resp.} $l^2$) computed with $P$ as time-dependent probability law on $(p, q)$.

All computations are carried out with $\epsilon_m m = 1$, $\epsilon_l=1$ and the same simple initial condition: $\psi^{-}(j=0,p,q) = 1$ if $(p,q)=(0,0)$ and $0$ elsewhere, $\psi^{+}(j=0,p ,q) = 0$ for all $(p,q)$. The only remaining free parameters are $\epsilon_A E$ and $\epsilon_A B$. 
As will now be discussed, DTQWs for which both $\epsilon_A E$ and $\epsilon_A B$ are much smaller than unity exhibit regimes which resemble continuous physics. DTQWs with larger values of $\epsilon_A E$ and $\epsilon_A B$ behave very differently, and can even localize.

\begin{figure}[h!]
 \includegraphics[width=8.5cm]{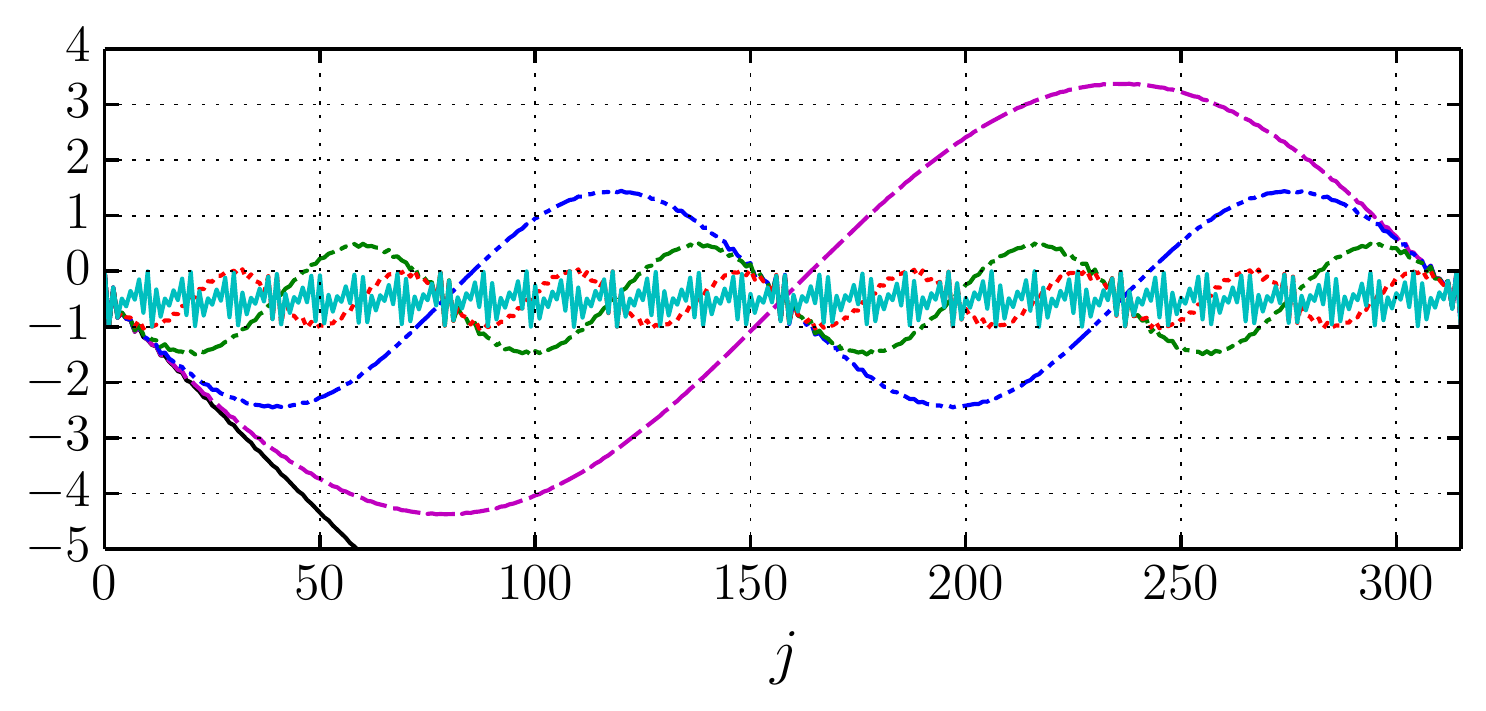}
        \vspace{-0.4cm}
       \caption{ Time evolution of the $p$-mean for $E$-walks with
$\epsilon_A E=0$ (black, solid), $0.02$ (magenta, dashed), $0.04$ (blue, dot-dot-dashed), $0.08$ (green, dot-dashed) $0.16$ (red, doted), $0.64$ (cyan, solid). The oscillating period is $T_{\mbox{\tiny{Bloch}}} = 2 \pi /(\epsilon_A E)$ with an error less than one lattice site. 
\label{fig:X_mean} }
\end{figure}

Figure \ref{fig:X_mean} shows the time evolution of the $p$-mean for several $E$-walks. 
For $\epsilon_A E=0$, the $p$-mean varies linearly with time.
This ballistic transport is typical of homogeneous DTQWs, {\sl i.e.} DTQWs whose coin operators do not depend on the spacetime point. Moreover, transport occurs towards negative values of $p$ only because the initial state has a vanishing $\psi^+$. For $\epsilon_A E\neq0$, the $p$-mean oscillates in time around the value $X^1=-0.5$ \cite{footnote}
with a period which coincides with the so-called Bloch period 
$
T_{\mbox{\tiny{Bloch}}}
=2\pi/(\epsilon_A E)$ with an error smaller than one time step. 
Bloch oscillations were first predicted by F. Bloch \cite{Bloch29} and C. Zener \cite{Zener34} for electrons moving in solids. They have been observed in $2D$ photonic lattices \cite{Trompeter06} and $1D$ electric DTQWs \cite{mesch13a,Regensburger}. As $\epsilon_A E$ reaches a sizeable fraction of $2\pi$,
$T_{\mbox{\tiny{Bloch}}}$ becomes of the order of a few time steps. Another oscillating mode with period of the order of one time step then appears and dominates the dynamics.


\vspace{-0.2cm}

\begin{figure}[h!]
  \includegraphics[width=8.7cm]{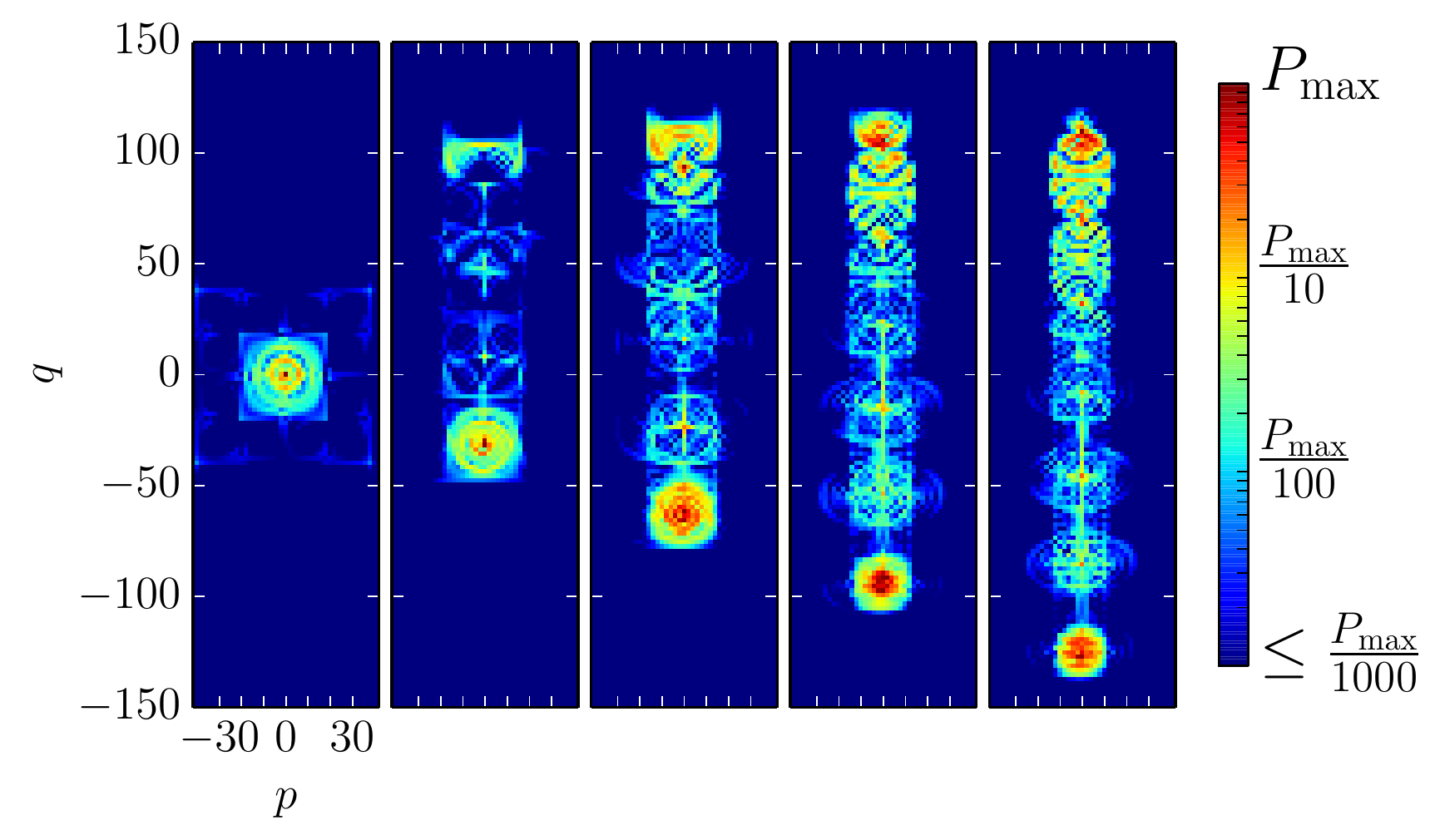}
   \vspace{-0.7cm}
        \caption{Probability density, at time $j=500$, for $EB$-walks with $\epsilon_A B=0.16$.
From left to right, $\epsilon_A E=0$, $0.01$, $0.02$, $0.03$, $0.04$, and $P_{\text{max}}=0.0943$, $0.0578$, $0.0209$, $0.0181$, $0.0178$. 
The bottom front corresponds essentially to the classical $\mathbf{E} \times \mathbf{B}$ drift.
 \label{fig:typical_drift} }
\end{figure}

\begin{figure}[h!]
        \includegraphics[width=8.5cm]{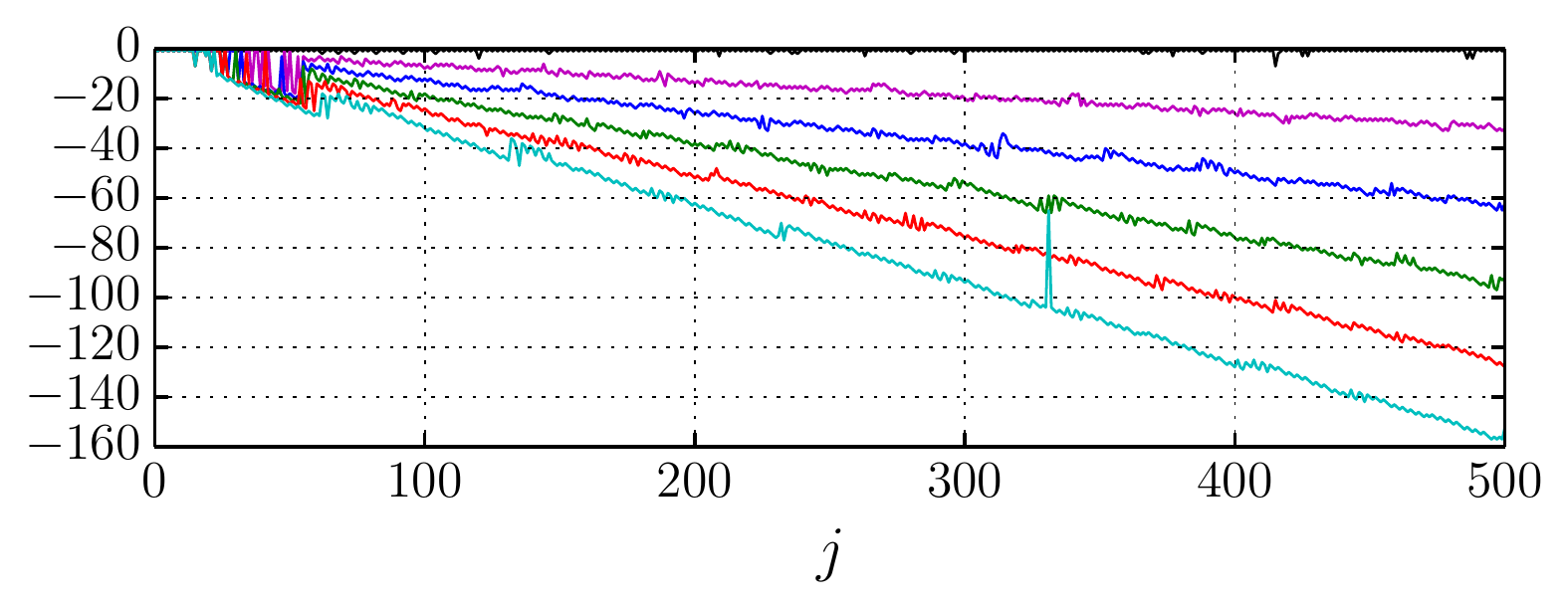}
        \vspace{-0.4cm}
        \caption{ Time evolution of the $q$-coordinate of the bottom-front local maximum of the probability density, for $EB$-walks with $B=0.16$.
From top to bottom, $E=0$ (black), $0.01$ (magenta), $0.02$ (blue), $0.03$ (green), $0.04$ (red) and $0.05 $ (cyan). This maximum propagates in the direction of $\mathbf{E} \times \mathbf{B}$ (up to small oscillations in the $p$ direction) and with speed $E/B$ up to a $1\%$ precision.
        \vspace{0cm} \label{fig:down_Y_maxloc} }
\end{figure}

Figure \ref{fig:typical_drift} displays the probability densities at time $j=500$ for several $EB$-walks with $\epsilon_A B=0.16$.
For $\epsilon_A E=0$ (left), the walker is quasi-confined around the origin, with a typical radius which slowly increases with the time $j$ and is, at each $j$, a decreasing function of $\epsilon_A B$ (data not shown, see \cite{ADmag16} for detail). When $\epsilon_A E\neq0$, the walker spreads in the $q$ direction, up and down. The bottom front propagates with a speed which coincides with $E/B$, as supported by Fig. \ref{fig:down_Y_maxloc}. This corresponds to the classical so-called `$\mathbf{E} \times \mathbf{B}$ drift' of a charged particle under crossed constant and uniform electric and magnetic fields (see, {\sl e.g.}, \cite{book_Jackson}). The roughly circularly symmetric `Landau profile' obtained for $\epsilon_A E=0$ seems to be transported at the drift velocity. The behaviour of the top front is counter intuitive from the classical perspective. The top-front spreads with a speed which seems independent of $\epsilon_A E$. A very similar behaviour has already been pointed out in \cite{KolovskyMantica14} for 
quantum particles moving under the influence of super-imposed electric and magnetic fields in a $2D$ periodic potential with tight-binding.

%

\begin{figure}[h!]
        \includegraphics[width=8.5cm]{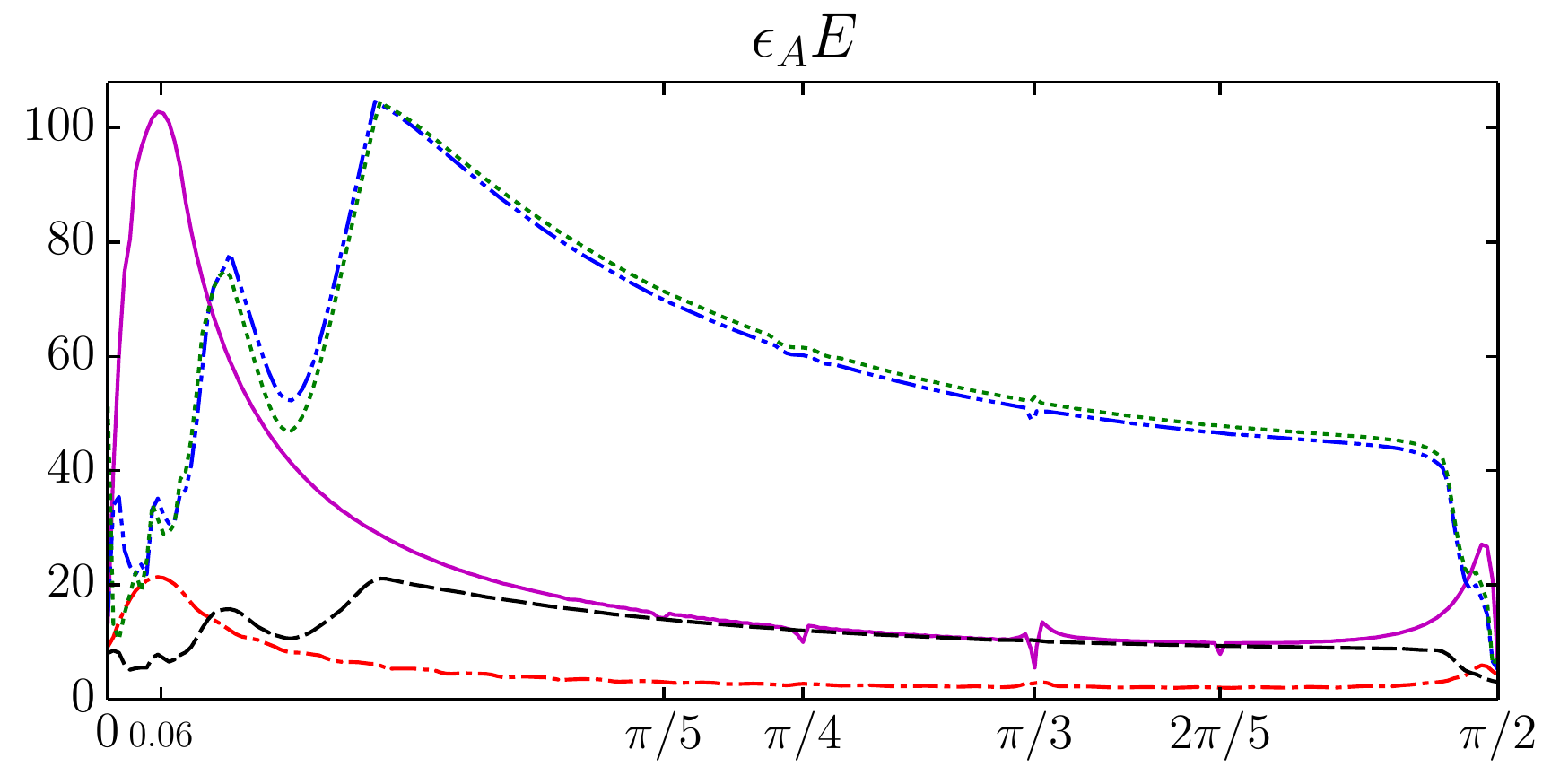}
        \vspace{-0.4cm}
        \caption{Evolution of the $q$-spread as a function of $\epsilon_A E$ for $EB$-walks with magnetic field (i) $\epsilon_A B=0.16$ at times $j=100$ (red, dot-dashed) and $j=500$ (magenta, solid) (ii) $\epsilon_A B=1$, at times $j=100$ (black, dashed) and $j=500$ (blue, dot-dot-dashed) (iii) $B=\pi/3\simeq 1.047$ (green, doted) at time $j=500$. 
        \vspace{0cm} \label{fig:versus_E} }
\end{figure}

\begin{figure}[h!]
        \includegraphics[width=8.5cm]{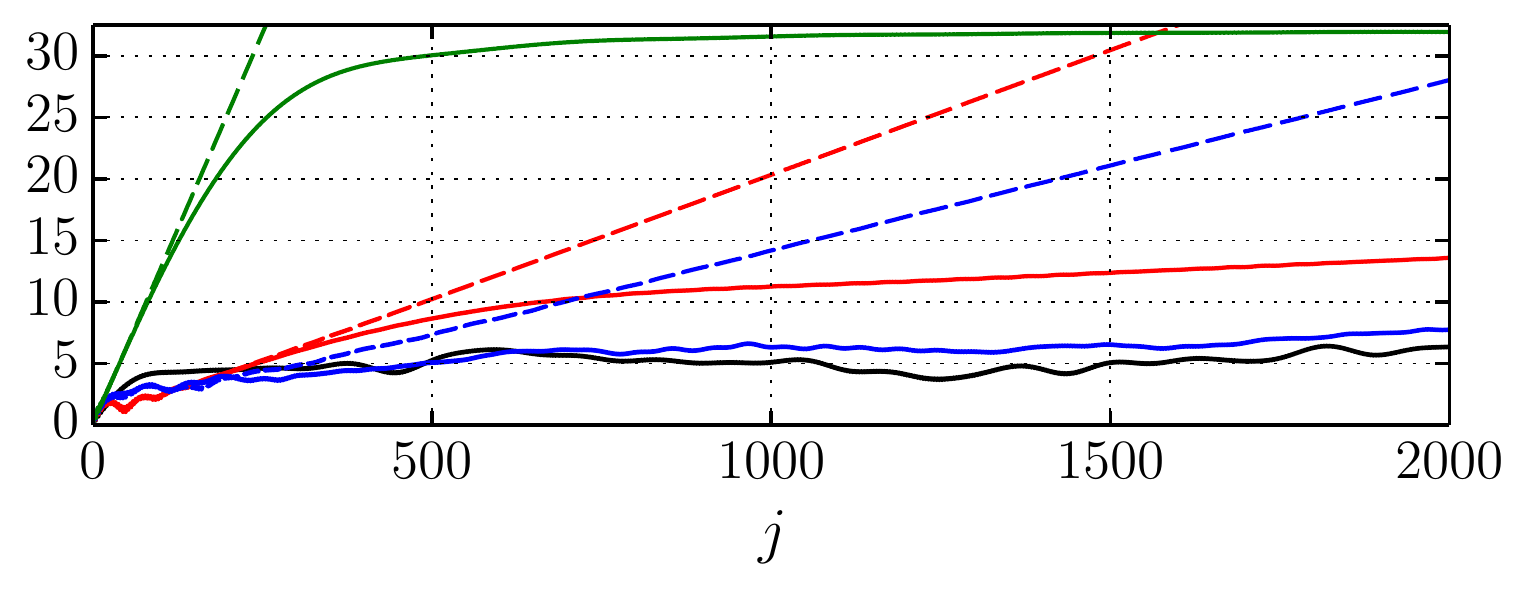}
        \vspace{-0.4cm}
        \caption{Time evolution of the $q$-spread for $EB$-walks with $\epsilon_A E=\pi/2$ and $\epsilon_A B = 0.16$ (black, solid), $\pi/4$ (red, dashed), $\pi/4+\varepsilon$ (red, solid), $\pi/3$ (blue, dashed), $\pi/3+\varepsilon$ (blue, solid), $\pi/2$ (green, dashed), $\pi/2+\varepsilon$ (green, solid), with $\varepsilon=0.04$.
        \vspace{0cm} \label{fig:versus_time_E_piover2} }
\end{figure}

Previous work on DTQWs coupled to electric or magnetic fields \cite{mesch13a,ced13,Yalcinkaya15} have shown that walks with field values which are rational multiples of $2 \pi$ (`rational fields') follow very peculiar dynamics.
Fig. \ref{fig:versus_E} displays the $q$-spread of $EB$-walks as a function of $\epsilon_A E$ at two times and different values of $\epsilon_A B$.
For $\epsilon_A B=0.16$, which is not a rational multiple of $2 \pi$, there is a weak $E$-field regime (from $\epsilon_A E=0$ to $\epsilon_A E\simeq0.06$) in which the $q$-spread increases essentially linearly with $\epsilon_A E$. 
This is the regime of Figures \ref{fig:typical_drift} and \ref{fig:down_Y_maxloc}. 
For $\epsilon_A E > 0.06$, the $q$-spread decreases considerably.
This weak $E$-field regime breaks down partially for $\epsilon_A B=1$ and completely for $\epsilon_A B=\pi/3$, while the $q$-spreading is essentially enhanced for strong values of $\epsilon_A E$. For $\epsilon_A E=\pi/2$ and values of $\epsilon_A B$ which are not rational multiples of $2 \pi$, the walk seems to be almost localized in $q$ (this is also the case in $p$ direction, data not shown). Fig. \ref{fig:versus_time_E_piover2} focuses on this apparent localization. In the long-time limit, the walker spreads ballistically for values of $\epsilon_A B$ which are rational multiples of $2 \pi$. This ballistic spreading is considerably reduced (quasi-localisation) for $\epsilon_A B=\pi/4+\varepsilon$ and $\pi/3+\varepsilon$, and the walk seems to really localise for $\epsilon_A B=\pi/2+\varepsilon$. The $p$-spread displays the same qualitative behaviours (data not shown).

\section{Conclusion and discussion} 

We have introduced a new family of $2D$ DTQWs which coincides, in the continuous limit, with the dynamics of a Dirac fermion coupled to arbitrary electromagnetic fields. The wavefunction of these DTQWs has two components and the DTQWs explore the $2D$ square lattice by advancing alternately in each of the orthogonal directions. Similar, albeit simpler $2D$ DTQWs have been discussed for example in \cite{DiFranco11, DMDmag13}.
We have shown that the new DTQWs introduced in this article possess an exact discrete local $U(1)$ gauge invariance, a discrete gauge-invariant conserved current and a discrete gauge-invariant electromagnetic field, and that field and current can be coupled by discrete generalizations of   Maxwell equations. We have also explored the behaviour of the DTQWs outside the continuous limit, under weak and strong fields. For weak fields, we have observed discrete versions of the Bloch oscillations and of the so-called ${\bf E} \times {\bf B}$ drift. We have also observed localization for some higher values of the fields.

The results of this article prove that DTQWs can be used to build full-fledged discrete gauge theories and that laboratory experiments based on quantum walks can, at least in principle, simulate these theories (see for example \cite{mesch13a} for a discussion of a quantum walk  experiment already carried out which simulates Dirac fermions coupled to $1D$ electric fields).
On the technical side, the construction we have presented should naturally be extended, not only to Maxwell electromagnetism in $4D$ spacetime, but also to other Yang-Mills gauge theories. Developing second-quantized versions of these discrete theories should also prove interesting.


A full comparison of possible discrete gauge theories based on DTQWs with the usual Lattice Gauge Theories (LGTs) \cite{Wilson74,book_Rothe} is beyond the scope of this article. Let us simply mention two differences. First, unlike the `U' parallel transporters in LGTs, gauge fields do not have to be added by hand to the DTQW dynamics, as the connection is already part of the basic definition of DTQWs and most DTQWs are by definition locally gauge invariant \cite{DMD14}. Second, the difference operators (discrete derivatives) which arise in conjunction with the local gauge invariance of DTQWs are more complicated than the usual finite difference operators used in lattice gauge theories. The mathematical properties of discrete gauge theories based on DTQWs are thus probably very different from the mathematical properties of LGTs.

Finally, DTQWs are useful in a much wider context than high energy or condensed matter physics. DTQWs are in particular universal building blocks of quantum algorithms \cite{Childs2009} and our results therefore have implications for quantum information. For example, the exploration of graphs by DTQWs could be influenced by creating discrete gauge fields on these graphs. Indeed, not only do gauge fields influence the transport of single DTQWs, but gauge theories provide a novel manner to implement interaction between DTQWs.

\end{document}